# State of the Art and Prospects for Quantum Computing

**M. I. Dyakonov**

*Laboratoire Charles Coulomb, Université Montpellier II, CNRS, France*

### 1. A grand challenge for the Millennium

Here is the title of one of many recent books devoted to quantum computing: "*Quantum Computation: A Grand Mathematical Challenge for the Twenty-First Century and the Millennium*", S. J. Lomonaco, Jr., ed. American Mathematical Society, Providence, Rhode Island (2002).

There are some points worth noticing: first, that we are supposed to work on quantum computing not only during the current century, but also for the next 900 years, and second, that the challenge we face is purely *mathematical*.

A thousand years is quite a lot of time, and it is very probable that the quantum computing business will be terminated much earlier. The curious reader who digs out this book in the year 3002, will look at it as we today would consider the title: "*Philosopher's Stone (Lapis Philosophorum): A Grand Challenge for the Eleventh Century and the Millennium*", Roman Philosophical Society, Pisa (1002).

Meanwhile, the activity in the field of quantum computing remains quite high with a tendency towards saturation, see Fig. 1. For how long will this rate of one article per day continue?

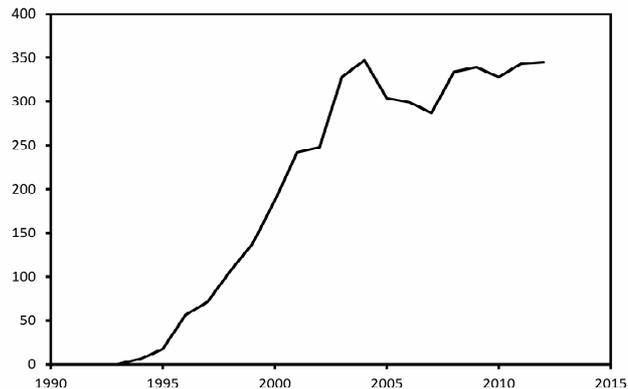

**Figure 1.** Number per year of articles in arXiv containing the words "quantum computation" in abstract

The existing literature can be divided into three unequal groups: Experiments, Proposals, and Theory. Roughly speaking, for each experimentally manipulated qubit there are 3 proposals and 30 proved theorems. Here, I will present a very brief review of the first two groups, as well as an analysis of the existing theoretical approach. Based on this analysis, the prospects for quantum computing remain uncertain.

**2. The general ideas of quantum computing**

The idea of quantum computing[1] is to store information in the values of $2^N$ complex amplitudes describing the wavefunction of $N$ two-level systems (*qubits*), and to process this information by applying unitary transformations (*quantum gates*), that change these amplitudes in a precise and controlled manner. The value of $N$ needed to have a useful machine is estimated to be $10^3$ or more. Note that even $2^{1000} \sim 10^{300}$ is much greater than the number of protons in the Universe.

Since the qubits are always subject to various types of noise, and the gates cannot be perfect, it is widely recognized that large scale, i.e. useful, quantum computation is impossible without implementing error correction. This means that the $10^{300}$ continuously changing quantum amplitudes of the grand wavefunction describing the state of the computer must closely follow the desired evolution imposed by the quantum algorithm. The random drift of these amplitudes caused by noise, gate inaccuracies, unwanted interactions, etc., should be efficiently suppressed.

It is not obvious at all that error correction can be done, even in principle, in an analog machine which state is described by at least $10^{300}$ continuous variables. Nevertheless, it is generally believed (for example, see Ref. 2) that the prescriptions for fault-tolerant quantum computation[3–6] using the technique of error-correction by encoding[7, 8] and concatenation (recursive encoding) give a solution to this problem. By active intervention, errors caused by noise and gate inaccuracies can be detected and corrected during the computation.

The so-called "threshold theorem"[9–11] says that, once the error per qubit per gate is below a certain value estimated as $10^{-6}$–$10^{-4}$, indefinitely long quantum computation becomes feasible, at a cost of substantially increasing the number of qubits needed. However, very luckily, the number of qubits increases *only polynomially* with the size of computation.

Thus, theorists claim that the problem of quantum error correction has been solved, at least in principle, so that physicists and engineers need only to work on finding the good candidates for qubits and on approaching the accuracy required by the threshold theorem: "As it turns out, it is possible to digitize quantum computations arbitrarily accurately, using relatively limited resources, by applying quantum error-correction strategies developed for this purpose"[12]. All the hopes for scalable quantum computing rely entirely on the threshold theorem.

### 3. ARDA Experts Panel roadmap

Ten years ago, in 2002, at the request of the Advanced Research and Development Activity (ARDA) agency of the United States government, a team of distinguished experts in quantum information established a roadmap for quantum computing[2] (updated in 2004), with the following five- and ten-year goals:

"by the year 2007, to
• encode a single qubit into the state of a logical qubit formed from several physical qubits
• perform repetitive error correction of the logical qubit, and
• transfer the state of the logical qubit into the state of another set of physical qubits with high fidelity, and

by the year 2012, to
• implement a concatenated quantum error-correcting code."

The 2007 goal requires "something on the order of ten physical qubits and multiple logic operations between them", while the 2012 goal "requires on the order of 50 physical qubits, exercises multiple logical qubits through the full range of operations required for fault-tolerant QC in order to perform a simple instance of a relevant quantum algorithm".

While a benevolent jury could consider the first two of the 2007 goals to be partly achieved by now (see Section 3), the expectations for the third 2007 goal, and especially for the 2012 goal, are wildly off the mark. So are some other expectations: "As larger-scale quantum computers are developed over the next five and ten years, quantum simulation is likely to continue to be the application for which quantum computers can give substantial improvements over classical computation".[2] Apparently, there is a qualitative difference between doing experiments with 5 qubits and with 50 qubits, and the reasons for this should be understood.

Referring to the threshold theorem, the Experts Panel also claimed: "It has been established, under certain assumptions, that if a threshold precision per gate operation could be achieved, quantum error correction would allow a quantum computer to compute indefinitely".

As we shall see, here, the key words are "*under certain assumptions*", even though other publications do not even mention any assumptions or restrictions at all, e.g. Ref. 12. However the Experts Panel did not address the crucial point of whether these assumptions can be realized in the physical world. We will take a closer look at this issue in Sects. 6 and 9, and this will help us to distinguish between what has been established and what has *not* been established.

It would be quite interesting to see an updated roadmap for the next decade.

## 4. Experiments

Experimental studies related to the idea of quantum computing make only a tiny part of the whole QC literature. They represent the *nec plus ultra* of the modern experimental technique, they are extremely difficult and inspire respect and admiration.

The goal of such proof-of-principle experiments is to show the possibility of manipulating small numbers of two-level quantum systems serving as qubits, to realize the basic quantum operations, as well as to demonstrate elements of the Shor's factoring algorithm[13] and error correction by encoding.[7, 8]

*Factoring 15.* The first experiment reporting factoring 15 by Shor was done by Vandersypen et al[14] using the liquid nuclear magnetic resonance (NMR) technique. All the gates were implemented by microwave pulses applied within about 1s, which is less than the nuclear decoherence time. The obtained NMR spectra corresponded very well to the predictions of Shor's procedure.

Lanyon et al[15] performed the same task in an optical experiment using the polarization of 4 photons, while Lucero et al[16] used Josephson qubits: "...we run a three-qubit compiled version of Shor's algorithm to factor the number 15, and successfully find the prime factors 48% of the time".

An aside about the "compiled version" of Shor's algorithm is in order. The full algorithm can factor a *k*-bit number using $72k^3$ elementary quantum gates; e.g., factoring 15 requires 4608 gates operating on 21 qubits.[17] Ref. 17 introduced a compiling technique which exploits properties of the number to be factored, allowing exploration of Shor's algorithm with a vastly reduced number of resources. One might say that this is a sort of (innocent) cheating: knowing in advance that 15=3×5, we can take some shortcuts, which would not be possible if the result were not known beforehand.

All the existing experimental testing of Shor's algorithm use this simplified approach. Recently, in a very remarkable work of Martin–López et al,[18] this approach allowed for the first time to factor 21 in an optical experiment, where an iterative procedure of recycling a single qubit was successfully implemented.

The demonstration of the *full* Shor's algorithm for factoring 15 is still *well beyond* the reach of experimental possibilities (see the prediction in Ref. 19).

As to the liquid NMR quantum computing technique, it seems that it should rather be regarded as a classical simulation of the quantum algorithm.[19] This was also pointed out in Refs. 15 and 20. "The ability to describe the correlations observed in NMR experiments in terms of classical correlations between the qubits casts doubt on the "quantumness" of the experiments".[20]

*Error correction.* Several recent outstanding experiments[21-23] (references to some earlier work can be found therein) demonstrated elementary error correction with 3 to 8 qubits in the line of the Experts Panel's 2007 goals.

Schindler et al[21] implement multiple quantum error correction cycles for phase-flip errors in a system of three trapped-ion qubits. Each cycle consisted of (i) encoding one qubit and two auxiliary (ancilla) qubits into an entangled state,

(ii) error incidence, (iii) detecting and correcting the error, and (iv) resetting the ancillas.

Reed et al[22] realized three-qubit quantum error correction with superconducting circuits by implementing the three-qubit Toffoli gate. They performed both bit- and phase-flip error corrections with fidelity around 80%.

Yao et al[23] reported the first experimental demonstration of "topological" error correction with a polarization-encoded eight-photon cluster state (see Refs. 23-25 for an explanation of the idea of topological cluster state computing). They have demonstrated that (i) if only one physical qubit suffers an error, the noisy qubit can be located and corrected, and (ii) if all qubits are simultaneously subjected to errors with equal probability, the effective error rate is significantly reduced by error correction.

The artificial errors introduced and then corrected in such experiments are *strong*: basically the state $|0\rangle$ is changed to $|1\rangle$. Using classical language, the error consists in reversing the direction of some vector (or rotating it by $90^o$).

However, the relevant errors, with which we are supposed to fight, are *weak*: the noise transforms the state $|0\rangle$ to $a|0\rangle+b|1\rangle$, with $|b|<<1$ (the vector is rotated by a small angle). Error correction consists in detecting the unwanted admixture of state $|1\rangle$ and eliminating this admixture. This task is especially difficult because in any experimental setup one will always have some uncontrolled admixture of state $|1\rangle$ from the start (see Section 8).

So, we will have to wait until somebody demonstrates the possibility to prepare his qubit in the state $|0\rangle$ with a precision of $10^{-6}$–$10^{-4}$, to detect an error of the same order of magnitude, and then correct it with the same precision. (In the eventual quantum computer such repetitive cycles are supposed to be done simultaneously with at least a thousand qubits).

## 5. Proposals. Quantum computing with...

The multitude of proposals of different ways to do quantum computing, as well as of various physical objects that can serve as qubits, is truly amazing. A simple list of such proposals would require the entire space allocated for this article. To show just the tip of the iceberg, below is a small number of randomly picked proposals (source: *arxiv.org*, references can be easily found there). For more information the reader is invited to google "quantum computing with".

*Quantum computing with:*
- non-deterministic gates
- bosonic atoms
- highly verified logical cluster states
- Pfaffian qubits
- hyperfine clock states
- four-dimensional photonic qudits
- quantum-dot cellular automata in dephasing-free subspace

- generalized binomial states
- 1D projector Hamiltonian
- quantum-dot spin qubits inside a cavity
- graphene nanoribbons
- alkaline earth atoms
- Jaynes-Cummings model
- doped silicon cavities
- Read-Rezayi states
- electron spin ensemble
- ultra narrow optical transition of ultracold neutral atoms in an optical lattice
- $p$-wave superfluid vortices
- railroad-switch local Hamiltonians
- global entangling gates
- semiconductor double-dot molecules
- decoherence-free qubits
- superqubits
- defects
- devices whose contents are never read
- alkaline-earth-metal atoms
- ionic Wigner crystals
- nanowire double quantum dots
- waveguide-linked optical cavities
- orbital angular momentum of a single photon
- probabilistic two-qubit gates
- non-deterministic gates
- small space bounds
- interaction on demand
- perpetually coupled qubits
- only one mobile quasiparticle
- moving quantum dots generated by surface acoustic waves
- para-hydrogen
- programmable connections between gates
- incoherent resources and quantum jumps
- nu = 5/2 fractional quantum Hall state
- spin ensemble coupled to a stripline cavity
- vibrationally excited molecules
  Kerr-nonlinear photonic crystals
- atoms in periodic potentials
- Heisenberg ABAB chain
- endohedral fullerenes
- harmonic oscillators
- ...

Isn't this wonderful? Apparently, there is nothing at all that is NOT suitable for quantum computing!

Let us consider just one of the proposals listed above, "quantum computing with the $v = 5/2$ fractional quantum Hall state".[26] The experimentally observed $v = 5/2$ Hall plateau is unlike all the others (for which the denominator of the filling factor $v$ is odd) and it does not fit into the composite fermion concept. Some people think that this is a manifestation of *anyons,* hypothetical particles intermediate between bosons and fermions that may exist in two dimensions. (See also the *Anyon theory of high $T_c$ superconductivity,*[27] now completely forgotten). Others think differently, and nobody really knows.

Hence the *obvious* proposal: to use these anyons for quantum computing. We must move them around one another producing complex topological structures, so that knot theory may be used. The great advantage is that topological structures are intrinsically protected against noise.[28]

The disadvantages are mostly on the practical side. The quantum Hall plateaus are observed in high magnetic field with a roughly 3×6mm sample immersed in liquid Helium. The sample has several wires attached, typically 6, to apply and measure voltages and currents. The number of 2D electrons in the sample is about $10^{10}$, and presumably it contains a similar number of anyons. So how are we supposed to create topological structures with these hypothetical anyons, just by applying voltages to the 6 (or, if one insists, even 6000) wires at our disposal?

However, this is an old story by now. Currently, it is supposed that the Majorana fermions,[29] rather than anyons, are responsible for the 5/2 Hall plateau. Consequently, it became obvious that quantum computing with Majorana fermions is extremely promising. The Majorana fermion theory of high $T_c$ superconductivity is likely to emerge in the near future.

The reader might have an immediate impulse to propose *topological insulators*, of which we have recently heard so much, for topological quantum computing, simply because both are "topological" (like "quantum dots for quantum computing"). Too late! This has been done already.[30]

A by-product of this frenetic activity is that every physical object has become a qubit, independently of whether it is regarded in the quantum computing context, or not:

> *Electron spin qubit*
> *Hole spin qubit*
> *Nuclear spin qubit*
> *Josephson superconducting qubit*
> *Cavity photon qubit*
> *Trapped ion qubit*
> *Para-Hydrogen qubit*
> *Heisenberg ABAB chain qubit*
> *etc.*

This looks pretty, like some modern poetry. So, instead of saying like in the good old days, "We are studying nuclear spin resonance", one should now say:

"We are studying decoherence of nuclear spin qubits", thus implying that this work is directly related to the big problems of the day.

## 6. Theory. Assumptions (axioms) underlying the threshold theorem

Like any theorem, this one also completely relies on a number of assumptions, considered as *axioms*:

1. Qubits can be prepared in the $|00000...00\rangle$ state. New qubits can be prepared on demand in the state $|0\rangle$,
2. The noise in qubits, gates, and measurements is uncorrelated in space and time,
3. No undesired action of gates on other qubits,
4. No systematic errors in gates, measurements, and qubit preparation,
5. No undesired interaction between qubits,
6. No "leakage" errors,
7. Massive parallelism: gates and measurements are applied simultaneously to many qubits,
and some others.

While clearly stated in the original work, the existence of these assumptions was largely ignored later, especially in presentations to the general public (Ref. 12 is one of many examples).

One would expect that the above assumptions, treated as axioms (i.e. as being *exact*), would undergo a close scrutiny to verify that they can be reasonably approached in the physical world. Moreover, the term "reasonably approached" should have been clarified by indicating with what precision each assumption should be fulfilled. So far, this has never been done (assumption 2 being an exception[31, 32]), if we do not count the rather naive responses provided in the early days of quantum error correction.[33–35]

It is quite normal for a theory to disregard small effects whose role can be considered as negligible. But not when one specifically deals with errors and error correction. A method for correcting *some* errors on the assumption that other (unavoidable) errors are *non-existent* is not acceptable, because it uses fictitious ideal elements as a kind of gold standard.[36]

## 7. Precision of continuous quantities and the basic axiom

Below are some trivial observations regarding manipulation and measurement of continuous variables. Suppose that we want to know the direction of a classical vector, like the compass needle. First, we never know exactly what our coordinate system is. We choose the *x, y, z* axes related to some physical objects with the *z* axis directed, say, towards the Polar Star, however neither this direction, nor the angles between our axes can be defined with an infinite precision. Second, the orientation of the compass needle with respect to the chosen coordinate system cannot be determined exactly.

So, when we say that our needle makes an angle $\theta = 45^\circ$ with the $z$ axis, we understand that $\cos\theta$ is not exactly equal to the irrational number $1/\sqrt{2}$, rather it is somewhere around this value within some interval determined by our ability to measure angles and other uncertainties. We also understand that we cannot manipulate our needles perfectly, that no two needles can ever point exactly in the same direction, and that consecutive measurements of the direction of the same needle will give somewhat different results.

In the physical world, continuous quantities can be neither measured nor manipulated exactly. In the spirit of the purely mathematical language of the quantum computing theory, this can be formulated in the form of the following

**Axiom 1**. No continuous quantity can have an exact value.
*Corollary.* No continuous quantity can be exactly equal to zero.

To a mathematician, this might sound absurd. Nevertheless, this is an unquestionable reality of the physical world we live in.[37, 38] Note that *discrete* quantities, like the number of students in a classroom or the number of transistors in the on-state, *can* be known exactly and *this* is what makes the great difference between the digital computer and an analog computer.

Axiom 1 is crucial whenever one deals with continuous variables. Thus if we devise some technical instruction, each step should contain an indication of the needed precision. Do not tell the engineer: "Make this angle $45^\circ$, and then my proposed vehicle will run as predicted, under the assumption that the road is flat". This makes no sense! Tell him instead: "Make this angle $45^\circ \pm 0.01^\circ$, and then my proposed vehicle will run as predicted, provided the roughness of the road does not exceed 10nm" (or 10cm, whatever the theory says). Only then the engineer will be in a position to decide whether this is possible or not.

All of this is quite obvious, and nobody is going to believe that even in a thousand years somebody will manage to make the angle exactly $45^\circ$ and provide an absolutely flat road to implement our invention.

## 8. Precision of quantum amplitudes

Apparently, things are not so obvious in the magic world of quantum mechanics. There is a widespread belief that the $|0\rangle$ and $|1\rangle$ states "in the computational basis" are something *absolute*, akin to the on/off states of an electrical switch, or of a transistor in a digital computer,[39] but with the advantage that one can use quantum superpositions of these states, see Fig. 2. It is sufficient to ask: "with respect to which axis do we have a spin-up state?" to see that there is a serious problem with such a point of view.

$$\Psi = A \cdot \boxed{\text{on}} + B \cdot \boxed{\text{off}}$$

**Figure 2.** The theorist's image of a qubit

It should be stressed once more that the coordinate system, and hence the computational basis, cannot be exactly defined, and this has nothing to do with quantum mechanics. Suppose that, again, we have chosen the *z* axis towards the Polar Star, and we measure the *z*-projection of the spin with a Stern-Gerlach beam-splitter. There will be inevitably some (unknown) error in the alignment of the magnetic field in our apparatus with the chosen direction. Thus, when we measure some quantum state and get (0), we never know exactly to what state the wavefunction has collapsed.

Presumably, it will collapse to the spin-down state with respect to the (not known exactly) direction of the magnetic field in our beam-splitter. However, with respect to the chosen *z* axis (this direction is not known exactly either) the wavefunction will always have the form $a|0\rangle + b|1\rangle$, where hopefully $|b| \ll 1$. Another measurement with a similar instrument or a consecutive measurement with the same instrument will give a different value of *b*.

Quite obviously, the unwanted admixture of the $|1\rangle$ state is an error that *cannot be corrected*, since (contrary to the assumption 1 above) we can never have the standard *exact $|0\rangle$* and $|1\rangle$ *states* to make the comparison.

Thus, with respect to the consequences of imperfections, the situation is quite similar to what we have in classical physics. The classical statement "the exact direction of a vector is unknown" is translated into quantum language as "there is an unknown admixture of unwanted states". The pure state $|0\rangle$ can never be achieved, just as a classical vector can never be made to point *exactly* in the *z* direction, and for the same reasons – after all quantum measurements and manipulations are done with classical instruments.

Clearly, the same applies to *any* desired state. Thus, when we contemplate the "cat state" $(|0000000\rangle + |1111111\rangle)/\sqrt{2}$, we should not take the $\sqrt{2}$ too seriously, and we should understand that *some* (maybe small) admixture of all other 126 possible states of 7 qubits must be necessarily present.

> *Exact quantum states do not exist. Some admixtures of*
> *all possible states to any desired state are unavoidable.*

This fundamental fact described by Axiom 1 (nothing can be *exactly* zero!) should be taken into account in any prescriptions for quantum error correction. Note, that the "digitization" of noise, which is the cornerstone of the existing error-correcting schemes, is based on the *opposite* assumption, that the exact $|0\rangle$ and $|1\rangle$ states *can* be prepared.

At first glance, it may seem that there *are* possibilities for achieving a desired state with an arbitrary precision. Indeed, using nails and glue, or a strong magnetic field, we can fix the compass needle so that it will not be subject to noise. We still cannot determine exactly the orientation of the needle with respect to our chosen coordinates, but we can take the needle's direction as the *z* axis. However: 1) we cannot align another fixed needle in exactly the same direction and 2) we cannot use fixed needles in an analog machine; to do this, they must be detached to allow for their free rotation.

Quite similarly, in the quantum case we can apply a strong enough magnetic field to our spin at a low enough temperature, and wait long enough for the relaxation processes to establish thermodynamic equilibrium. Apparently, we will then achieve a spin-down $|0\rangle$ state with any desired accuracy (provided there is *no interaction* with other spins in our system, which is hardly possible).

However, "spin-down" refers to the (unknown exactly) direction of the magnetic field at the spin location. Because of the inevitable inhomogeneity of the magnetic field, we cannot use the direction of the field at the spin location to define the computational basis, since other spins within the same apparatus will be oriented slightly differently. Moreover, if we want to manipulate this spin, we must either switch off the magnetic field (during this process the spin state will necessarily change in an uncontrolled manner), or apply a resonant ac field at the spin precession frequency, making the spin levels degenerate in the rotating frame. The high precision acquired in equilibrium will be immediately lost.

Likewise, an atom at room temperature may be with high accuracy considered to be in its ground state. Atoms at different locations will be always subject to some fields and interactions, which mix the textbook ground and excited states. Also, such an atom is not yet a two-level system. In order for it to become a qubit, we must apply a resonant optical field, which will couple the ground state with an excited state. The accuracy of the obtained states will depend on the precision of amplitudes, frequencies, and duration of optical pulses. This precision might be quite sufficient for many applications, but certainly it can never be *infinite*. Thus, Axiom 1 applies to *all* continuous variables, quantum amplitudes included.

## 9. The fundamental trouble with the error-correction theory

The trouble lies in not respecting Axiom 1, i.e. in assuming that all the numerous assumptions (axioms) behind the threshold theorem are fulfilled *exactly*. This is not possible, not in our world.[37]

Of course, *if* the assumptions underlying the threshold theorem were approached with a *high enough precision*, the prescriptions for error-correction could indeed work. So, the real question is: *what* is the required precision with which each assumption should be fulfilled to make scalable quantum computing feasible? How small should be the undesired, but unavoidable: interaction between qubits, influence of gates on other qubits,[40] systematic errors of gates and measurements,[34] leakage errors, random and systematic errors in preparation of the initial $|0\rangle$ states, etc.?

Quite surprisingly, not only is there no answer to these most crucial questions in the existing literature, but they have never even been seriously discussed! Had this problem been realized, the threshold theorem would not be formulated in terms of "error per qubit per gate" only, but also by indicating the required *precision* with which various assumptions should be fulfilled. Obviously, this gap must be filled, and the required precision for each assumption should be specified.

Until this is done, one can only speculate about the final outcome of such a research. The optimistic prognosis would be that some additional threshold values $\varepsilon_1$, $\varepsilon_2$... for corresponding precisions will be established, and that these values will be shown neither to depend on the size of the computation nor to be extremely small. In this case, the dream of factoring large numbers by Shor's algorithm might become reality in some distant future.

The pessimistic view, which I share, is that the required precision must increase with the size of computation, and this would be a farewell to quantum computing.

Classical physics gives us some enlightening examples regarding attempts to impose a prescribed evolution on quite simple continuous systems. Consider some number of hard balls in a box. At $t=0$ all the balls are on the left side and have some initial velocities. We let the system run for some time, and at $t=t_0$ we simultaneously reverse all the velocities. Classical mechanics tells us that at $t=2t_0$ the balls will return to their initial positions in the left side of the box. Will this ever happen in reality, or even in computer simulations?

The known answer is: Yes, *provided* the precision of the velocity inversion is exponential in the number of collisions during the time $2t_0$. If there is some slight noise during the whole process, it should be exponentially small too. Thus, if there are only 10 collisions, our task is difficult but it still might be accomplished. But if one needs 1000 collisions, it becomes impossible, not because Newton Laws are wrong, but rather because the final state is strongly unstable against very small variations of the initial conditions and very small perturbations.

This classical example is not directly relevant to the quantum case (see Ref. 41 for the relation between classical and quantum chaos). However it might give a hint to explain why, although some beautiful and very difficult experiments with small numbers of qubits have been done (see Section 3 for recent results with 3 to 8 qubits), the goal of implementing a concatenated quantum error-correcting code with 50 qubits (set by the ARDA Experts Panel[2] for the year 2012!) is still nowhere in sight.

To summarize, I reiterate my main points:

1. The hopes for scalable quantum computing are founded entirely on the "threshold theorem": once the error per qubit per gate is below a certain value, indefinitely long computations are possible.

2. The mathematical proof of the threshold theorem heavily relies on a number of assumptions supposed to be fulfilled *exactly*, as axioms.

3. In the physical world nothing can be exact when one deals with continuous quantities. For example, it is not possible to have *zero* interaction between qubits, it can only be *small*.

4. Since the basic assumptions cannot be fulfilled *exactly*, the question is: What is the required *precision* with which each assumption should be fulfilled?

5. Until this crucial question is answered, the prospects of scalable quantum computing will remain uncertain.

**10. Mathematics and physical reality**

Besides having an enormous intrinsic value, mathematics is indispensable for understanding the physical world, as well as for all practical human activities. However, there are many ways of abusing mathematics.

For example, you want to discuss the very complex phenomenon of *love*. If you are a poet or a sexologist or, better still, if you have some experience of your own, you may have chances to say something reasonable. But if the only thing you have up your sleeve is to write the Hamiltonian of the couple as $H = H_1 + H_2 + V$, with $V$ being a sum of products of operators belonging to the subspaces 1 and 2, and then, *under certain assumptions*, you prove some theorems, quite obviously your rigorous results will be both wrong and irrelevant. And the reason is that you simply have no idea even about $H_1$, and still less about $H_2$, not to mention $V$. As a consequence, whatever your assumptions are, they are groundless.[42]

Another example, nearer to our subject, can be found in Jaroslav Hašek's masterpiece.[43] The good soldier Švejk spends some time in a madhouse, where one of the professors among the patients tries to convince everybody that "Inside the terrestrial globe there is another globe of a much greater diameter".

In fact, that mad professor's claim is consistent with the well-known fact that, in certain metric spaces, a ball of a bigger radius may be properly contained in one of a smaller radius. Another (perfectly rigorous) result, which he also anticipated in a way, is the famous Banach-Tarski paradox that was discovered 10 years later;[44] see the very clear article in Wikipedia.[45]

A version of the Banach-Tarski theorem states that, given a small ball and a huge ball in the usual 3D Euclidean space, either one can be partitioned into pieces and then reassembled into the other. Note that the number of pieces is *finite*, and the reassembly process consists in moving them around, using only translations and rotations (but no stretching). Thus, as Wikipedia puts it, "a pea can be chopped up and reassembled into the Sun".

The interested reader must learn about metric spaces, nonmeasurable sets, and other mathematical technicalities in order to understand how it is possible that a rigorously proved theorem contradicts our common sense, and whether we should revise our common sense accordingly. (The short answer is *no*, because our common sense is based on the structure of the surrounding physical world, while the axioms behind the theorem are not.)

Now, imagine some society on another planet, where an army of scientific journalists not familiar with the technicalities cites the distinguished experts, affirming (quite correctly, *in some sense*) that *it has been proved* that one can build a full-scale skyscraper on the basis of its 1000:1 model without using any additional material, and more importantly, that the same applies to a tank or a submarine. Scientists in top-secret labs are developing the technical procedures, and many promising materials have been already proposed, such as endohedral fullerenes, ionic Wigner crystals, and some others.

That's what is happening in some places of our planet with respect to the threshold theorem: "The theory of fault-tolerant quantum computation establishes that a noisy quantum computer can simulate an ideal quantum computer accurately. In particular, the quantum accuracy threshold theorem asserts that an arbitrarily long quantum computation can be executed reliably, provided that the noise afflicting the computer's hardware is weaker than a certain critical value, the *accuracy threshold*".[46] In other words, the possibility of scalable quantum computing has been rigorously proved. As we have seen, this is not true.

No theorem can be proved without a set of axioms on which it relies. It is not the mathematician's concern whether his axioms describe the physical reality correctly or not. However, this should be the *main* concern for those who want to apply the theorem to the physical world. This question lies outside mathematics and the only way to solve it is to consult experiment and the "engineers". They will ask some hard questions, because "decoherence-free subspaces" and "approaching with arbitrary precision any unitary transformation on *n* qubits by an appropriate number of gates from the universal set" is not in their vocabulary.

Dorit Aharonov, one of the authors of the threshold theorem, wrote in 1998: "In a sense, the question of noisy quantum computation is theoretically closed. But a question still ponders our minds: Are the assumptions on the noise correct?" [47]

And indeed, the question is closed *in the sense* that, based on a number of axioms, the theorem is proved. However, if by "noise" we mean *all* sort of uncertainties and undesired disturbances inevitably occurring in reality, then the assumptions on the noise are *not* correct, because *some* unavoidable noise is assumed to be absent, and *some* quantum states are assumed to be *exact*.

In this *other sense*, the question remains wide open. Until somebody specifies the required precision with which various assumptions should be approached, the prospects of scalable quantum computing will be far from clear.

## 11. More powerful in doing what?

Suppose that, in spite of the above arguments, a quantum computer will be built one day. Why should anyone need it?

Fifteen years ago, John Preskill wrote:[48] "But suppose I could buy a truly powerful quantum computer off the shelf today – what would I do with it? I don't know, but it appears that I will have plenty of time to think about it! My gut feeling is that if powerful quantum computers were available, we would somehow think of many clever ways to use them".

At the same time, Andrew Steane put it somewhat differently[1]: "The idea of 'Quantum Computing' has fired many imaginations simply because the words themselves suggest something strange but powerful, as if the physicists have come up with a second revolution in information processing to herald the next millenium.[49] This is a false impression. ... If large quantum computers are ever made, they will be used to address just those special tasks which benefit from quantum information processing."

So, after 15 years of thinking, what are those clever ways and those special tasks? Apparently, just like fifteen years ago, there are only two of them:[50] factoring very large numbers by Shor's algorithm and simulation of quantum systems, which was the original idea of Feynman[51] who started the whole field 30 years ago. Let us discuss the utility of these two potential applications.

*Factoring*. Some (but not all) of existing cryptographic systems depend on the difficulty of factoring very large numbers (products of two or several very large primes). This means that Shor's algorithm could break cryptography codes. In turn, this implies the need for quantum cryptography.

Thus we encounter the following logic: 1) We will build quantum computers, then we will be able to know our enemies' secrets, and this is good. 2) However, after some time our enemies will also manage to build their own quantum computer, then *they* will be able to know *our* secrets, this would be very bad. 3) Anticipating this occurrence, we should develop the methods of quantum cryptography now, even before the first quantum computer is built.

So, it appears that the justification for implementing large scale Shor's algorithm is solely in opening the need for quantum cryptography, because otherwise the existing cryptography methods are quite safe. This logic would be understandable, from a certain point of view, had it not been for the existence of *several* cryptography systems, that *do not* depend on factoring, and *cannot* be broken by a quantum computer, at least no methods of doing this are known.[52]

Thus there seems to be no practical sense in performing factoring with a quantum computer. In my opinion, this does not diminish the importance of Shor's outstanding theoretical achievement.[13]

*Simulating quantum systems*. In 1996 Lloyd[53] proved that Feynman's conjecture,[51] that quantum computers can be programmed to simulate any local quantum system, is correct. Some people believe that, because of this, quantum computing will make a revolution in physics, chemistry, and biology.

The last statement seems too strong, since so far numerical calculations and simulations, while useful, were never of primary importance for these fields. However, obviously many theorists would like to have a device that could efficiently solve quantum problems of strongly interacting particles, when no simplifying approximations can be justified. Since the dimension of the Hilbert space increases exponentially with the number of particles *N*, making numerical calculations for *N*=50 is usually impossible and one has to be content with small values of *N*, up to 10 or 15, and extrapolate the results to the limit *N*=∞. This is not always easy, so it would be really nice if it were possible to do such calculations, say, up to $N$=1000 (see the recent review[54] for existing ideas and algorithms for quantum simulation).

In summary, it appears that simulation of strongly interacting quantum systems is the only meaningful possible application of the hypothetical quantum computer. Certainly, this would be an interesting and useful achievement, but hardly revolutionary, unless we understand this term in some very narrow sense.

**12. Quantum computing as a sociological problem**

I believe that, in spite of appearances, the quantum computing story is nearing its end, not because somebody proves that it is impossible, but rather because 20 years is a typical lifetime of any big bubble in science, because too many unfounded promises have been made, because people get tired and annoyed by almost daily announcements of new "breakthroughs",[55] because all the tenure positions in quantum computing are already occupied, and because the proponents are growing older and less zealous, while the younger generation seeks for something new.

The brilliant works of Feynman,[51] Deutsch,[56] Shor[13], and some others, will certainly remain for a long time because new and audacious ideas are always valuable, whether they lead to practical results, or not. However, this does not apply to the major part of the huge QC literature.

In fact, quantum computing is not so much a scientific, as a sociological problem which has expanded out of all proportion due to the US system of funding scientific research (which is now being copied all over the world). While having some positive sides, this system is unstable against spontaneous formation of bubbles and mafia-like structures. It pushes the average researcher to wild exaggerations on the border of fraud and sometimes beyond. Also, it is much easier to understand the workings of the funding system, than the workings of Nature, and these two skills only rarely come together.

The QC story says a lot about human nature, the scientific community, and the society as a whole, so it deserves profound psycho-sociological studies, which should begin right now, while the main actors are still alive and can be questioned.

A somewhat similar story can be traced back to the 13$^{th}$ century when Nasreddin Hodja[57] made a proposal to teach his donkey to read and obtained a 10-year grant from the local Sultan. For his first report he put breadcrumbs between the pages of a big book, and demonstrated the donkey turning the pages with his hoofs. This was a promising first step in the right direction.

Nasreddin was a wise but simple man, so when asked by friends how he hopes to accomplish his goal, he answered: "My dear fellows, before ten years are up, either I will die or the Sultan will die. Or else, the donkey will die."

Had he the modern degree of sophistication, he could say, first, that there is no theorem forbidding donkeys to read. And, since this does not contradict any known fundamental principles, the failure to achieve this goal would reveal new laws of Nature.[58] So, it is a win-win strategy: either the donkey learns to read, or new laws will be discovered.

Second, he could say that his research may, with some modifications, be generalized to other animals, like goats and sheep, as well as to insects, like ants, gnats, and flies, and this will have a tremendous potential for improving national security: these beasts could easily cross the enemy lines, read the secret plans, and report them back to us.

The modern version of these ideas is this love-song for military sponsors: "The transistors in our classical computers are becoming smaller and smaller, approaching the atomic scale. The functioning of future devices will be governed by quantum laws. However, quantum behavior cannot be efficiently simulated by digital computers. Hence, the enormous power of quantum computers will help us to design the future quantum technology."

This may look good to a project manager, or in some science digest magazine, but for anyone who understands something about simulation, quantum laws, transistors, and atoms, this does not make any sense at all.

The saga of quantum computing is waiting for a deep sociological analysis, and some lessons for the future should be learnt from this fascinating adventure.

**Acknowledgement.** I thank Konstantin Dyakonov for very useful discussions and suggestions.

computer will resemble a huge microwave oven. It will be a real challenge to exclude the unwanted action of gates on other qubits.

41. M. C. Gutzwiller, Chaos in classical and quantum mechanics, Springer, New-York (1990)

42. To put it bluntly, one should refrain from proving theorems about systems and phenomena of which one has no profound understanding.

43. Jaroslav Hašek, "The Fateful Adventures of the Good Soldier Švejk During the World War", translated by Zenny K. Sadlon, SAMIZDAT (2007); amazon.com/Fateful-Adventures-Soldier-Svejk-During/dp/1585004286

44. S. Banach, A. Tarski, "Sur la décomposition des ensembles de points en parties respectivement congruents", *Fundamenta Mathematicae*, **6**, 244 (1924)

45. www.wikipedia.org/wiki/Banach_Tarski

46. P. Aliferis, D. Gottesman, and J. Preskill, "Accuracy threshold for postselected quantum computation", *Quantum Inf. Comput.* **8**, 181 (2008); arXiv:quant-ph/0703264

47. D. Aharonov, "Quantum computation", in: *Annual Reviews of Computational Physics*, D Stauffer, ed., World Scientific, **VI**, p. 259 (1999); arXiv:quant-ph/981203

48. J. Preskill, "Quantum Computing: Pro and Con", *Proc. Roy. Soc. Lond.* **A454**, 469 (1998); arXiv:quant-ph/9705032

49. This is exactly what is being told to the general public even today.

50. Quantum algorithms that provide (with an *ideal* quantum computer!) only polynomial speed-up compared to digital computing, like the Grover algorithm, became obsolete due to the polynomial slow-down imposed by error correction.

51. R. P. Feynman, "Simulating physics with computers", *Int. J. Theor. Phys.* **21**, 467 (1982); "Quantum mechanical computers", *Found. Phys*. **16**, 507 (1986)

52. D. J. Bernstein, "Introduction to Post-Quantum Cryptography", in: *Post-quantum cryptography*, D. J. Bernstein, J. Buchmann, E. Dahmen, eds., Springer, Berlin (2009)

53. S. Lloyd, "Universal quantum simulators", *Science* **273**, 1073 (1996)

54. K. L. Brown, W. J. Munro, and V. M. Kendon, "Using quantum computers for quantum simulation", *Entropy* **12**, 2268 (2010); arXiv:1004.5528

55. When a spin relaxation time of 10ns is measured, instead of 1ns previously, this is heralded as another breakthrough on the way to quantum computing.

56. D. Deutsch, "Quantum theory, the Church-Turing principle and the universal quantum computer". *Proceedings of the Royal Society of London; Series A, Mathematical and Physical Sciences,* **400,** 97 (1985)

57. Nasreddin Hodja was a populist philosopher and a wise man believed to have lived around 13[th] century during the Seljuq dynasty and remembered for his funny stories and anecdotes. The International Nasreddin Hodja fest is celebrated annually in July in Aksehir, Turkey every year.

58. This kind of reasoning is quite common in our days: "Because there are no known fundamental obstacles to such scalability, it has been suggested that failure to achieve it would reveal new physics"[12].